\documentclass[sigconf]{acmart}
\usepackage{tabularx}
\AtBeginDocument{%
  }

\usepackage{balance}

\usepackage{listings}
\usepackage{xcolor}
\usepackage{colortbl}
\usepackage{pifont}
\usepackage{soul}

\usepackage{adjustbox}
\usepackage{booktabs}

\lstdefinestyle{customStyle}{
  basicstyle=\ttfamily,
  keywordstyle=[1]\bfseries, 
  keywordstyle=[2]\underline, 
  morekeywords=[1]{Commit}, 
  morekeywords=[2]{speedup, performance} 
}

\usepackage{color}
\usepackage{tabularray}
\usepackage{hhline}
\usepackage{caption}
\usepackage{graphicx}

\usepackage[utf8]{inputenc}
\usepackage{tcolorbox}

\newcommand{\tool}[0]{\mbox{\textit{\textsc{PerfCurator}}}}
\newcommand{\modelkd}[0]{\mbox{\textit{\textsc{PcBERT-KD}}}}
\newcommand{\modelhs}[0]{\mbox{\textit{\textsc{PcBERT-HS}}}}

\setcopyright{acmlicensed}
\copyrightyear{2018}
\acmYear{2018}
\acmDOI{XXXXXXX.XXXXXXX}

\acmConference[Conference acronym 'XX]{Make sure to enter the correct
  conference title from your rights confirmation emai}{June 03--05,
  2018}{Woodstock, NY}
\acmISBN{978-1-4503-XXXX-X/18/06}

\renewcommand\footnotetextcopyrightpermission[1]{} 

\usepackage{tikz}
\usepackage{amsmath}

\usepackage{adjustbox}
\usepackage{multirow}

\begin{document}

\title{\tool{}: Curating a large-scale dataset of performance bug-related commits from public repositories}

\author{Md Abul Kalam Azad}
\authornote{Both authors contributed equally to this research.}
\email{akazad@umich.edu}
\affiliation{%
  \institution{University of Michigan Dearborn}
  \city{Dearborn}
  \state{Michigan}
  \country{USA}
}

\author{Manoj Alexender}
\authornotemark[1]
\email{manojale@umich.edu}
\affiliation{%
  \institution{University of Michigan Dearborn}
  \city{Dearborn}
  \state{Michigan}
  \country{USA}
}

\author{Matthew Alexender}
\email{mattalex@umich.edu}
\affiliation{%
  \institution{University of Michigan Dearborn}
  \city{Dearborn}
  \state{Michigan}
  \country{USA}
}

\author{Syed Salauddin Mohammad Tariq}
\email{ssmtariq@umich.edu}
\affiliation{%
  \institution{University of Michigan Dearborn}
  \city{Dearborn}
  \state{Michigan}
  \country{USA}
}

\author{Foyzul Hassan}
\email{foyzul@umich.edu}
\affiliation{%
  \institution{University of Michigan Dearborn}
  \city{Dearborn}
  \state{Michigan}
  \country{USA}
}

\author{Probir Roy}
\email{probirr@umich.edu}
\affiliation{%
  \institution{University of Michigan Dearborn}
  \city{Dearborn}
  \state{Michigan}
  \country{USA}
}

\renewcommand{\shortauthors}{Azad et al.}
\newcommand{\Azad}[1]{\textcolor{blue}{{ [ #1]}}}
\newcommand{\Manoj}[1]{\textcolor{green}{{ [ #1]}}}
\newcommand{\Foyzul}[1]{\textcolor{red}{{ [ #1]}}}

\begin{abstract}
Performance bugs challenge software development, degrading performance and wasting computational resources. Software developers invest substantial effort in addressing these issues. Curating these performance bugs can offer valuable insights to the software engineering research community, aiding in developing new mitigation strategies. However, there is no large-scale open-source performance bugs dataset available. To bridge this gap, we propose \tool{}, a repository miner that collects performance bug-related commits at scale. \tool{} employs \modelkd{}, a 125M parameter BERT model trained to classify performance bug-related commits. Our evaluation shows \modelkd{} achieves accuracy comparable to 7 billion parameter LLMs but with significantly lower computational overhead, enabling cost-effective deployment on CPU clusters. Utilizing \modelkd{} as the core component, we deployed \tool{} on a 50-node CPU cluster to mine GitHub repositories. This extensive mining operation resulted in the construction of a large-scale dataset comprising 114K performance bug-fix commits in Python, 217.9K in C++, and 76.6K in Java. Our results demonstrate that this large-scale dataset significantly enhances the effectiveness of data-driven performance bug detection systems.
\end{abstract}


\maketitle


\section{Introduction}
Performance bugs are a notorious challenge that degrade software performance and waste computational resources~\cite{Jin2012,Nistor2013}. These bugs are associated with reduced end-user satisfaction, increased development and maintenance costs, and diminished revenues~\cite{Liu2014,Nistor2013}. Due to their pervasive nature, performance bugs will continue to emerge as software and hardware evolve and new areas of computing emerge. Despite their persistence, identifying and fixing performance bugs in software remains a significantly challenging task for developers~\cite{kalam_azad_empirical_2023}.

Detecting and fixing performance bugs requires an in-depth understanding of various software and hardware components, including algorithms, concurrency, libraries, programming languages, runtime, and hardware architectures. However, prior research~\cite{kalam_azad_empirical_2023} has shown that only a handful of developers possess the extensive experience needed to understand and resolve these performance issues. To democratize performance-efficient coding practices, significant innovations in tooling support are necessary to provide novice developers with deep insights and code recommendations.

\begin{lstlisting}[caption={List of performance and non-performance commit descriptions. Keywords are underlined.}, label={lst:patch_description}, frame=single, style=customStyle, 
escapechar=\#, captionpos=b]
Commit-1:28-30% improvement in cuda 
vs opencl speedup for bilateral filter #\ding{52}#
Commit-2:Add padding to avoid false-sharing #\ding{52}#
Commit-3:Add comment referencing Chrome 
performance bug for Array.splice #\ding{56}#
\end{lstlisting}

To achieve this goal, it is essential for the performance tool research community to be informed about current end-user challenges. This knowledge will enable the development of new performance analysis techniques that address the ever-evolving software and hardware landscape. However, there is currently no central repository of performance bugs to curate this knowledge. In contrast, the software security community maintains a vulnerability database that informs researchers about common vulnerabilities and drives innovation in detection tools~\cite{cve}. We envision that a common repository for performance bugs would drive the innovation of performance techniques and inform developers of best practices. Moreover, the recent excitement in the software engineering community regarding language models and their effectiveness in generating or recommending performance-efficient code will require a large-scale, high-quality performance dataset.

Software code repositories are an abundant source of performance bugs and bug-fix efforts. Prior research~\cite{kalam_azad_empirical_2023, Nusrat2021, Nistor2013,zaman2012qualitative,han2016empirical} has manually analyzed code commits in these repositories to identify various categories of performance bugs. However, curating a large dataset of performance bugs from code commits is challenging. The current method for identifying performance-related commits is rudimentary: earlier studies have relied on keywords (such as performance, speed up, accelerate, fast, efficient, optimize, etc.) to flag relevant commits. This approach leads to numerous false positives and false negatives, requiring human effort to manually investigate and label each commit. Listing~\ref{lst:patch_description} demonstrates examples of performance and non-performance-related commit messages. For instance, the keyword \textit{performance} misclassifies commit-3, whereas commit-2 goes undetected with the current list of performance-related keywords. Given the wide range of performance topics and the limited knowledge of known performance bugs, creating new keyword lists and manually analyzing them is a daunting process that demands significant expertise and effort while still missing a substantial number of performance bugs. A scalable approach that does not rely on manual labeling is needed to effectively curate performance bugs from software repositories.

The quest leads to the research question, (\textbf{RQ1}) \textit{Are advanced Natural Language Processing (NLP) techniques, such as Large Language Models (LLMs), better at identifying performance commits in the wild?} Through experimental evaluation, we demonstrate that recent large language models, such as the 7 billion parameter Mistral-7B~\cite{jiang2023mistral}, are proficient at detecting performance commits with high accuracy. However, Mistral-7B and its quantized variants are large models that require significant computational resources for inference. Due to these high computational demands, leveraging them in practical settings, such as CI/CD pipelines and large-scale data collection, becomes challenging.

\begin{figure}[!t]
\centering
    \includegraphics[width=\linewidth]{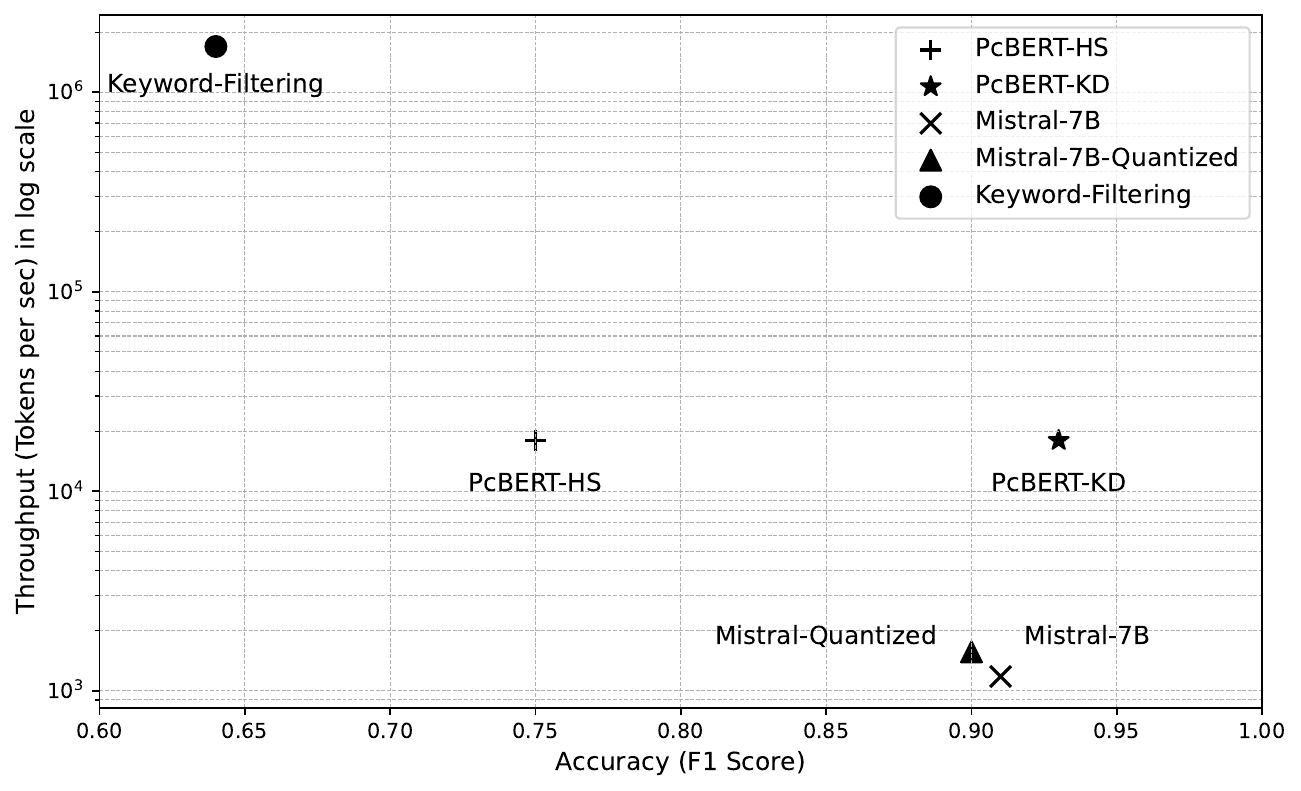}
    \caption{Throughput vs Accuracy for various approaches of commit message classification}
    \label{fig:ThroughputVsAccuracy}
\end{figure}

To find a practical solution, we further investigate our next research question, (\textbf{RQ2}) \textit{Can we train a lightweight transformer, such as a 125M parameter Bidirectional Encoder Representations from Transformers (BERT) model, to classify performance commits with high accuracy?} We investigate two approaches: 
\begin{itemize}
    \item \textbf{Heuristic Supervision:} We train a 125M parameter BERT model, \textit{\modelhs{}}, using this weak supervision learning technique and find that it performs significantly better than keyword-based classification. However, \textit{\modelhs{}} accuracy still lags behind that of the Mistral-7B models.
    \item \textbf{Knowledge Distillation:} We leverage the Mistral-7B model to train a 125M parameter transformer model, \textit{\modelkd{}}, using this supervised learning technique to classify performance commits. Empirical evaluation shows that \textit{\modelkd{}} performs as well as the Mistral-7B models.
\end{itemize} 

We then investigate (\textbf{RQ3}) \textit{How computationally intensive are the smaller 125M-parameter transformer models in comparison to the larger Mistral 7B models?}. We benchmark on both GPU and CPU servers, revealing that the 125M-parameter transformer models require significantly fewer computational resources compared to the larger 7B parameter models and their quantized variants. Figure~\ref{fig:ThroughputVsAccuracy} plots the log-scale throughput versus accuracy of all the approaches on a NVIDIA GeForce RTX 4090 GPU workstation.

We implement \tool{}, a repository mining tool designed to collect a large-scale dataset of performance-related commits from GitHub repositories. At its core, \tool{} leverages the \textit{\modelkd{}} model to identify commits related to performance bug fixes. We deploy \tool{} on a 50 node CPU cluster and collect 114K, 217.9K, and 76.6K performance commits written in Python, C++, and Java, respectively. To understand the quality of the collected dataset, we further investigate (\textbf{RQ4}) \textit{What is the distribution of performance commits across different performance categories as identified by the \textit{\modelkd{}} model?}. We find that the \textit{\modelkd{}} model can identify a wide range of performance commits, including memory optimization, elimination of unnecessary computations, algorithmic optimization, and API misuse.

\begin{figure}[!t]
\centering
     \includegraphics[width=0.90\linewidth]{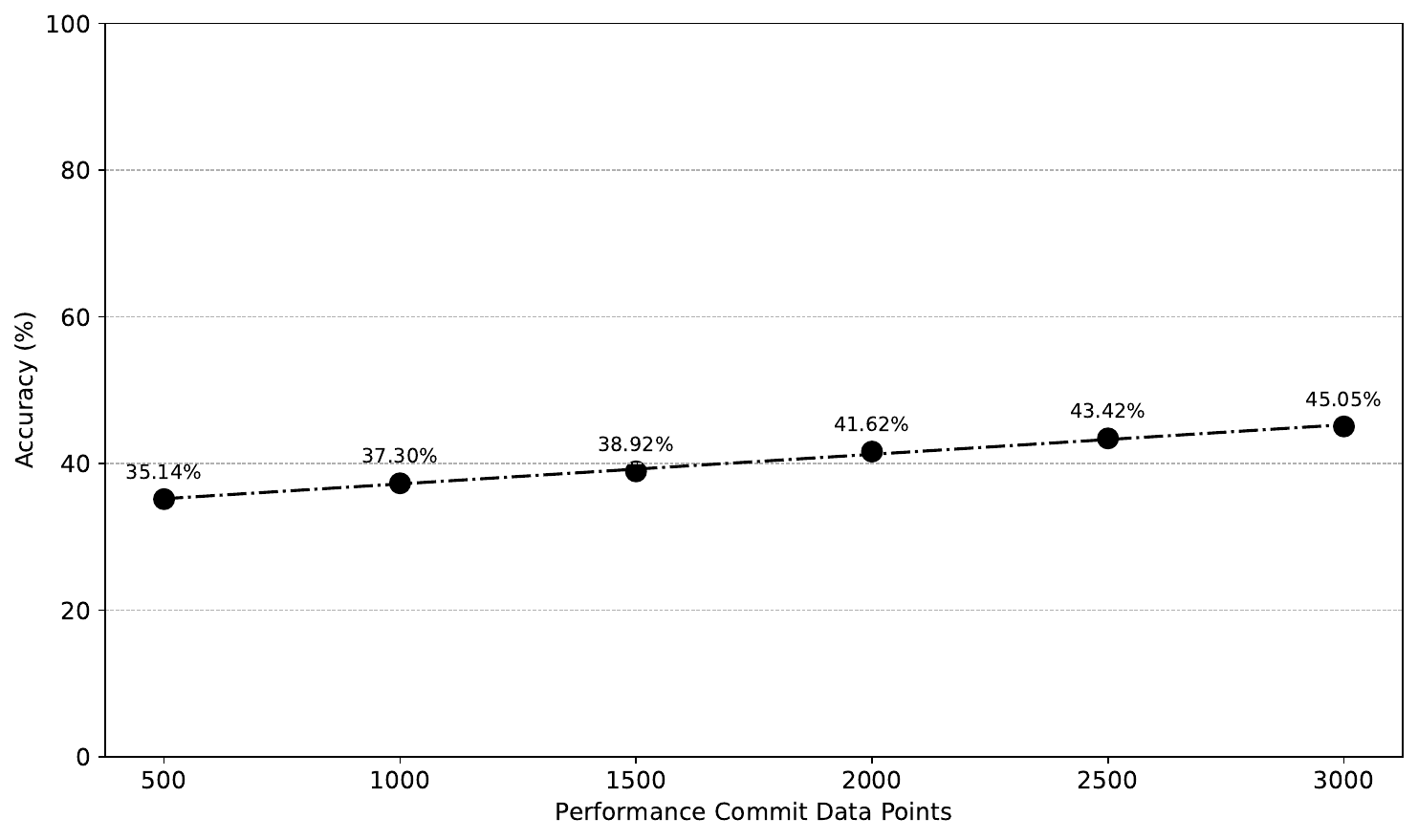}
    \caption{Accuracy of API misuse detection improved with increased performance commit data points collected by \tool{}.}
    \label{fig:accuracy_scability}
\end{figure}

Finally, to address (\textbf{RQ5}) \textit{How effective is the large-scale dataset in detecting inefficient coding practices}, we apply a data-driven API misuse detection tool. Through empirical evaluation, we demonstrate that data-driven API misuse detection tools perform significantly better with a large-scale dataset. Figure~\ref{fig:accuracy_scability} shows that the accuracy of API misuse detection improves as the performance commit data points increase. 

\paragraph{\textbf{Paper Contributions}}
This paper makes the following contributions:
\begin{itemize}
\item We are the first to develop a language model-based technique to classify performance-related code commits \textbf{at scale}.
\item We perform thorough benchmarking, confirming that the proposed approach provides high accuracy at a low computational cost.
\item Leveraging the proposed classifier, we have developed a repository mining tool and constructed a large-scale dataset comprising 114K, 217.9K, and 76.6K performance commits written in Python, C++, and Java, respectively. This dataset encompasses a wide range of performance improvements.
\item We empirically demonstrate that the large dataset collected using \tool{} can significantly enhance data-driven performance bug detection systems. This advancement mitigates the previously required extensive human effort, thereby optimizing the data collection process and enhancing the efficacy of performance bug study.
\end{itemize}




\section{Related Work}
\paragraph{\textbf{Software performance bugs}} Performance bugs in software are often caused by developers implementing inefficient code sequences during development. These code inefficiencies can further lead to significant resource wastage. While performance bugs often require relatively simple source code changes that can significantly speed up software, they are difficult to discover because they do not exhibit fail-stop symptoms like functional bugs. Therefore, many studies have explored the characteristics of performance bugs that occur in software across various domains, such as traditional software~\cite{jin2012understanding,yang2018not,zaman_qualitative_2012,Nistor2013,liu_characterizing_2014}, high-performance computing~\cite{kalam_azad_empirical_2023}, machine learning~\cite{long_reporting_2022,cao_characterizing_2021}, blockchain~\cite{wan_bug_2017}, and autonomous vehicles~\cite{garcia_comprehensive_2020}. While current approaches to studying performance bugs require significant manual effort, the scalable approach of the proposed \tool{} will facilitate an in-depth exploration of the unexplored areas of software performance challenges with ease.

\paragraph{\textbf{Performance monitoring and analysis tools}}
A substantial body of research has focused on the development of performance analysis tools for various applications. Tools such as \texttt{GProf}~\cite{graham1982gprof}, \texttt{OProfile}~\cite{levon2006oprofile}, \texttt{HPCToolKit}~\cite{adhianto2010hpctoolkit}, \texttt{VTune Profiler}~\cite{vtune},  \texttt{DTrace}~\cite{cantrill2004dynamic}, and \texttt{TAU Performance System}~\cite{shende2006tau} are designed to pinpoint code regions with significant execution times, thereby assisting developers in performance optimization efforts. Moreover, a variety of specialized performance analysis techniques are available to identify particular resource inefficiencies in code. These include memory profilers~\cite{288540}, GPU profilers~\cite{nvidia_nvprof, amd_rocprofiler}, network profilers~\cite{opentracing, zhang2022crisp}, and tools for addressing concurrency challenges~\cite{serebryany2009threadsanitizer}. Despite the extensive support provided by these tools, developers continue to face challenges related to performance inefficiencies. The emergence of new computing domains necessitates the development of specialized tools to meet these evolving challenges. We envision that the dataset collected by \tool{} will offer valuable insights for the performance tool community, highlighting gaps in existing tools in meeting developers' challenges in writing efficient code and fostering innovation.

\begin{figure}[]
  \centering
  \includegraphics[width=\linewidth]{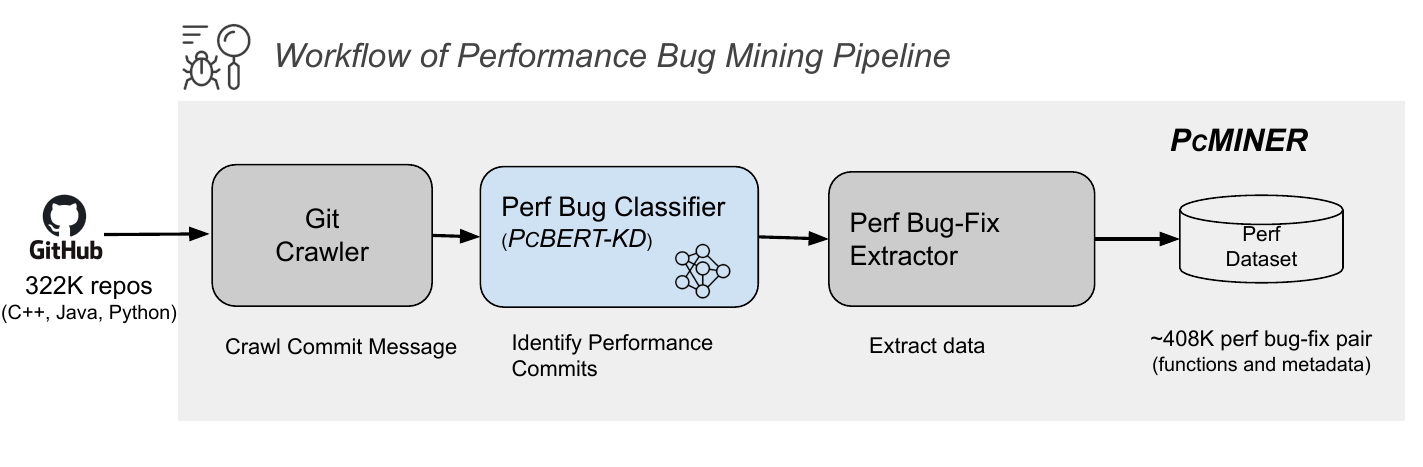}
  \caption{Workflow of \tool{} pipeline}
  \label{fig:pcMinerArch}
\end{figure}

\section{Methodology}

Figure~\ref{fig:pcMinerArch} provides an overview of \tool{}. The \tool{} consists of two primary components: \textcircled{\small{1}} the performance bug-related commit classifier, and \textcircled{\small{2}} the repository miner. This paper explores three \textit{BERT}-based classifiers and compare performance against a keyword-filtering baseline. In this section, we first discuss the construction of ground truth to validate the results of these classifiers (Section~\ref{groundtruth}). We then introduce our baseline; Keyword-Filtering (Section~\ref{keyfilter}) followed by a Large Language Model-based classification approach (Section~\ref{LLM_intro}). We then introduce two novel classifiers, \modelhs{} and \modelkd{} (Section~\ref{section:methodology}). Finally, we discuss the implementation of \tool{} for mining large-scale performance bug-related commits (Section~\ref{sec:miner}).
\subsection{Ground Truth Construction}
\label{groundtruth}

To evaluate the accuracy of the various classifiers, we at first curate a dataset of performance-related commits by mining GitHub repositories and manually labeling them. We develop a Python script that utilizes the GitHub API~\cite{github_api} and PyDriller~\cite{spadini2018pydriller} to mine these repositories. For data mining, we at first select 191,246 Python repositories with more than 20 stars to ensure they meet a minimum standard of quality. Through our random sampling, we find that less than 1\% of the commits are related to performance bugs. Due to high imbalance of performance-related commits, manually identifying performance bug fix commits from these repositories are challenging. Furthermore, keyword based filtering may fail to identify a large number of performance commits resulting in biased dataset.

To address the challenge, we employ a stratified random sampling strategy~\cite{sarndal2003model}. Our initial analysis reveals that large language models like Mistral 7B can identify performance commits that keyword-based searches overlook. We employ Mistral 7B to examine Python repositories, categorizing the commits into two classes: performance and non-performance commits. We then randomly sample each classes and collect equal number of samples for manual analysis. This process yields 100 performance commits and 100 non-performance commits.
\begin{figure}[!htbp]
\centering
\includegraphics[width=\linewidth]{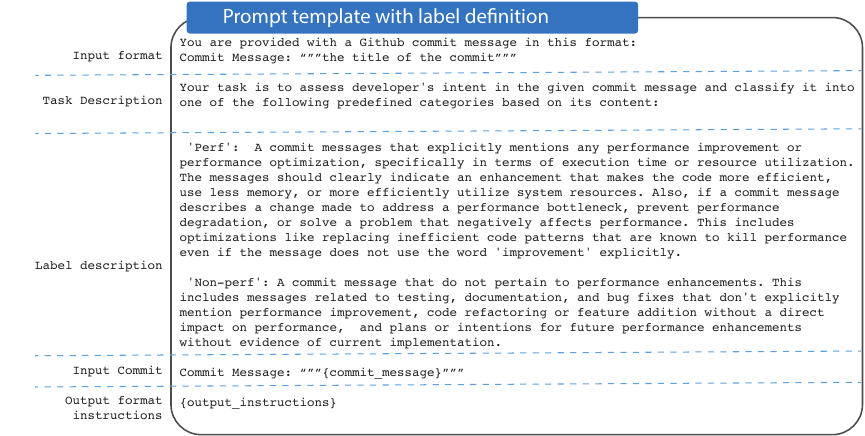}
\caption{Prompt template. In our \texttt{zero-shot} experimental setting, we provide label descriptions. (with temperature = 0)}
\label{fig:prompt_classification}
\end{figure}

Since Python is a high-level managed language, its applications are less likely to optimize architecture-specific, low-level code. Our initial analysis of the collected dataset supports this observation. To gather a diverse range of performance bug fix commits, we further examine 23 High Performance Computing (HPC) repositories referenced in prior literature~\cite{azad2023empirical}. However, since the authors collected the reported HPC performance bugs using keyword-based filtering, directly leveraging the curated commits might introduce bias into our dataset. Therefore, we apply the aforementioned stratified random sampling strategy to the repository list followed by manual analysis. This process results in another set of 150 performance commits and 150 non-performance commits.

Three authors independently label the dataset as performance and non-performance. After manual labeling, we compare the labels to identify any conflicts. To resolve these conflicts, we determine the final label through voting. We then calculate the Fleiss' kappa value to measure inter-rater agreement. The score for classifying the commits into performance and non-performance categories is 0.698, indicating a substantial level of agreement among the raters.

\subsection{Keyword-Filtering: The Baseline}
\label{keyfilter}

For our comparative analysis, we establish keyword-filtering as the baseline. We construct a list of keywords associated with performance bug-related commits, as identified in prior literature~\cite{kalam_azad_empirical_2023,Nusrat2021,chen_inferring_2019}. This list is compiled from various qualitative studies on performance bugs across multiple domains, including Cloud Computing, High-Performance Computing, Mobile Applications, and Deep Learning. The comprehensive nature of this keyword list allows for the identification of performance bug-related commits across a wide range of application domains. The list comprises 30 unique keywords, including \textit{performance, speed up, accelerate, fast, efficient, optimize}, among others.

\subsection{RQ1: Large Language Models for Classification}
\label{LLM_intro}

\paragraph{\textbf{Motivation}} 
Our initial observations indicate that keyword-filtering is inadequate for identifying domain-specific performance bug-related commits. The dynamic and evolving nature of computation presents significant challenges in maintaining and updating the keyword list. Moreover, as demonstrated in Listing~\ref{lst:patch_description}, keywords frequently mislabel commits. Manually identifying the false positives is an impractical and unsustainable approach.

\begin{table*}[]
\centering
\scriptsize
\vspace{-0.1cm}
\begin{tabular}{@{}ccccccc@{}}
\hline 
\textbf{Model} & \textbf{\#~Params.} &  \textbf{\#~Layers}  & \textbf{Window Size}  & \textbf{Pre-training Set} & \textbf{Architecture} & \textbf{Generative} \\ \hline \hline
\textbf{Mistral-7B} \cite{jiang_mistral_2023} & 7B & 32 & 32K & Open Web & Encoder-Decoder & Generative \\  
\textbf{RoBERTa} \cite{liu2019roberta} & 125M & 12 & 512 & Wikipedia, BookCorpus \cite{liu2019roberta} & Encoder-only & Non-generative \\ \hline 
\end{tabular}
\caption{Description of the Models Used in the Study}
\label{table:dataset}
\end{table*}

\paragraph{\textbf{Mistral-7B}} In the pursuit of developing robust classifiers capable of minimizing false positives and understanding the nuances of performance commits, we leverage recent advancements in large language models. In this study, we selected \textit{Mistral-7B}~\cite{jiang_mistral_2023} as the ideal candidate for this task. Table~\ref{table:dataset} details the features of the Mistral-7B model. Mistral-7B, an open-source model, demonstrates high performance and efficiency in natural language understanding. Benchmark studies~\cite{Mistral7BAnnouncement2023} indicate that Mistral-7B significantly outperforms Llama 2 13B across all metrics due to its distinctive features, such as Grouped-query attention (GQA)~\cite{ainslie2023gqa} for faster inference and Sliding Window Attention (SWA)~\cite{beltagy2020longformer} to handle longer sequences at smaller cost. The open-source nature of the model (Apache 2.0) allows for execution both locally and in the cloud without restrictions. In this study, we further evaluate \textit{Mistral-7B-AWQ}~\cite{Mistral7B-AWQ2023}, a quantized version of Mistral-7B designed to enhance efficiency and reduce computational resources while maintaining performance.
\paragraph{\textbf{Prompt Template}}
Figure~\ref{fig:prompt_classification} illustrates the prompt template used to classify a commit message as either performance-related or non-performance-related. The prompt includes a description of the classification task and definitions for each class label. Given our zero-shot experimental setting, we manually craft and verify the label explanations to guide the model in performing the classification. To ensure focused and deterministic outcomes, minimizing the risk of incorrect or unpredictable results, we set the temperature to 0~\cite{MistralAPI}.


\paragraph{\textbf{Observation}}
We rigorously evaluate the Mistral-7B and Mistral-7B-AWQ models using the ground truth dataset, with detailed results presented in section~\ref{section:RQ3}. The Mistral-7B model achieves a significantly higher F1-score of 0.92, compared to the baseline keyword-filtering method, which achieved an F1-score of 0.64. However, both Mistral-7B and its quantized variant incur substantial computational overhead, resulting in a 1,436$\times$ and 1,080$\times$ slowdown in throughput, respectively, compared to keyword-based filtering.

\paragraph{\textbf{Implication}}
Due to the significant computational cost of the Mistral 7B model, deploying it in practical settings to identify large-scale performance bug-related commits is challenging. Additionally, we anticipate integrating these detection tools into a performance bug curator, which will continuously monitor new commits and collect new knowledge to keep the community informed about performance bug trends. To overcome this challenge, the remainder of this paper explores alternatives to large models to minimize the computational overhead of performance bug related commit detection.

\subsection{RQ2: Training Small Language Models}
\label{section:methodology}
\label{subsection:overview}
\paragraph{\textbf{Overview}}
Pre-trained BERT-based neural network models have demonstrated significant effectiveness in various software engineering tasks. Unlike larger language models (LLMs), BERT (Bidirectional Encoder Representations from Transformers) employs a bidirectional transformer architecture, enabling it to consider context from both directions in a sequence. This bidirectional processing allows BERT to achieve a deeper understanding of linguistic and syntactic structures. Furthermore, these smaller language models (LMs) efficiently process input sequences and encode the structures of both natural and programming languages. Consequently, BERT-based models frequently outperform traditional approaches in numerous software engineering tasks, including code summarization~\cite{ahmad2020transformer}, bug explanation~\cite{mahbub_defectors_2023}, and recommending bug fixes~\cite{tufano2019empirical}.


Building on these advancements, we pose the research question: \emph{Can we fine-tune smaller transformers to classify performance commits with high accuracy?} To investigate this question, we explore two approaches for training transformers: \textcircled{\small{1}} heuristic supervision-based training and \textcircled{\small{2}} knowledge distillation-based training. These approaches resulted in two transformer models: \modelhs{} and \modelkd{}, respectively.

\subsubsection{\textbf{\modelhs{}: A Transformer Trained with Heuristic Supervision}} 
\begin{figure}[htbp]
    \centering
    \scalebox{0.5}{\input{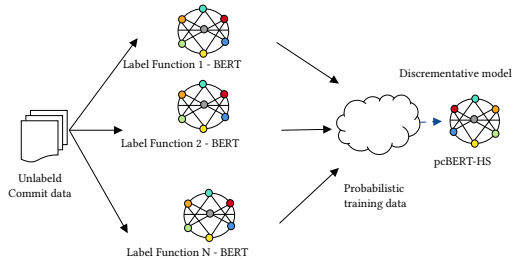}}
    \caption{Heuristic supervision-based learning process. Each LFs are 125M parameter BERT model, trained on a dataset labeled by a regular expression pattern}
    \label{fig:hs_pipeline}
\end{figure}

\paragraph{\textbf{Motivation}}
Due to the lack of extensive hand-labeled commit data necessary for training a robust classifier, we initially aim to develop our classification model using the heuristic-supervision technique as outlined in \cite{ratner_accelerating_nodate}. Heuristic supervision is a weak supervision strategy that leverages multiple noisy heuristics from domain experts to produce probabilistic labels, facilitating the creation of labeled datasets without ground truth annotations. This technique scales effectively to large datasets and has shown state-of-the-art performance in various domains~\cite{ratner2017snorkel, varma2019learning}. Figure~\ref{fig:hs_pipeline} illustrates the steps involved in heuristic supervision.

\paragraph{\textbf{Overview}} \textcircled{\small 1} We first identify and list the heuristics of performance and non-performance commit messages using regular expression patterns. \textcircled{\small 2} We then encode these heuristics in labeling functions (LFs)~\cite{ratner2017snorkel}. These LFs are leveraged to annotate the unlabeled dataset with soft labels. \textcircled{\small 3} We train a final 125-million parameter BERT model, \modelhs{}, on these soft labels under a weak supervision framework. \emph{The goal of this weak supervision training is to learn the implicit features of performance-related topics from the large commit dataset.}

\paragraph{\textbf{Dataset construction for training and validation}} To create our training datasets, we randomly select approximately 8 million commit messages from the GitHub Archive~\cite{grigorik2012github} using Google BigQuery~\cite{google_bigquery_github}. We use 7 million commit messages as the training dataset and the remaining 1 million for the validation set. From the 7 million training dataset, we further split it into 1 million messages to train the labeling functions and the remaining 6 million to extract soft labels for the final \modelhs{} training. For validation dataset, we select 2000 randomly sampled commits from the 1 million dataset. At least 60\% of these commits match our regular expression patterns. We then randomly sample equally from the matched performance and non-performance class to obtain 250 performance and 250 non-performance commits. Three authors independently label the data, achieving a Fleiss' Kappa value of 0.698.

\paragraph{\textbf{\textcircled{\small 1} Heuristic construction}}
We build a robust set of heuristics to classify performance and non-performance commits, following the strategy outlined in \cite{nema2022analyzing}. Initially, we prepare a list of bi-grams through manual exploration and prior knowledge. We then craft regular expression heuristics to capture diverse patterns of performance pitfalls from this bi-gram list. This list is iteratively expanded and refined through manual evaluation. Analyzing the words and expressions in commit messages helps us refine these regular expression patterns, effectively capturing variations in performance and non-performance-related commit descriptions. Our final set includes 71 regular expression patterns, encompassing both performance and non-performance classifications.

\paragraph{\textbf{\textcircled{\small 2} Labeling functions (LFs)}} 
We leverage labeling functions (LFs) \cite{ratner2017snorkel} to embed heuristics and label the dataset. 
A labeling function \( f_i \) takes a data point \( x \) as input and produces an integer label \( y \) corresponding to a class (performance or non-performance). Additionally, a labeling function can abstain from voting by outputting -1. Formally, 
\[ 
f_i(x) = 
\begin{cases} 
1 & \text{if performance-related} \\
0 & \text{if non-performance-related} \\
-1 & \text{if abstain}
\end{cases}
\]
However, we do not directly use the identified regular expressions as labeling functions. Instead, we utilize these regular expressions to inform and construct the LFs, enhancing their effectiveness in accurately labeling the dataset.


Regular expressions only match specific patterns and lack the ability to understand context or semantics beyond the predefined patterns. Due to this inherent limitation, we train individual 125M parameter BERT models based on the initial heuristics identified by each regular expression. We utilize the 1 million commits set aside for training labeling functions. 

We first label this 1 million commit dataset by each of the regular expressions and use them to train a corresponding BERT model. 
For each performance-related regular expression \( r_i \), we define a dataset \( D_i \) consisting of commits matching \( r_i \):
\[ D_i = \{ x_j \mid r_i(x_j) = 1 \text{ or } r_i(x_j) = -1 \} \]
Each \( D_i \) is used to train a BERT model \( \text{BERT}_i \), with parameters \( \theta_i \), to predict the likelihood of a commit being performance-related:
\[ \text{BERT}_i(x) = P(y = 1 \mid x; \theta_i) \]
Similarly, we train BERT models for non-performance-related regular expressions. Each non-performance regular expression \( r'_i \) labels the dataset \( D'_i \):
\[ D'_i = \{ x_j \mid r'_i(x_j) = 0 \text{ or } r'_i(x_j) = -1 \} \]
Each \( D'_i \) is used to train a BERT model \( \text{BERT}'_i \), with parameters \( \theta'_i \), to predict the likelihood of a commit being non-performance-related:
\[ \text{BERT}'_i(x) = P(y = 0 \mid x; \theta'_i) \]

From the identified 71 regular expressions, we train 71 individual BERT models. These individual BERT models are then used as labeling functions to annotate the 6 million unlabeled commit dataset with soft labels. We finally build a comprehensive label model \( \mathcal{M} \) which combines the outputs of individual LFs to generate probabilistic labels \( \hat{y} \) for the 6 million commit messages. For fair training, we construct a balanced dataset consisting of 200K performance commits and 200K non-performance commits from the 6 million labeled dataset. Table~\ref{table:sizeOFdatasets} details the dataset.


\paragraph{\textbf{\textcircled{\small 3} Training \modelhs{}}}
We then use this balanced dataset and their soft labels to train \modelhs{}, a final BERT model with 125M parameters. This probability-based training allows the \modelhs{} model to generalize better, as it learns from the nuanced signals provided by the aggregated soft labels. Consequently, the resulting transformer model is capable of capturing more complex patterns and providing more accurate classifications. This improved performance extends even to data points that do not perfectly match any single regular expression pattern.
\paragraph{\textbf{Observation}}
We conduct a rigorous evaluation of \modelhs{} using the ground truth dataset, with results detailed in Section~\ref{section:RQ3}. The \modelhs{} demonstrates a significantly higher F1-score of 0.75 compared to the baseline keyword-filtering method, indicating its success in learning new implicit features. Nonetheless, the performance of \modelhs{} remains below that of the Mistral-7B models.


\subsubsection{\textbf{\modelkd{}: A Transformer Trained by Knowledge Distillation}} 

\paragraph{\textbf{Motivation}}
The goal of this approach is leveraging the extensive knowledge embedded in large language models to develop a smaller, yet efficient model using the technique of knowledge distillation. Knowledge distillation facilitates the transfer of complex patterns and implicit features captured by a large, pre-trained teacher model to a smaller student model, thereby achieving high performance with reduced computational requirements~\cite{hinton2015distilling}. This approach mitigates the challenges associated with deploying large models, such as increased inference latency and substantial computational demands, making it suitable for real-time applications~\cite{gou2021knowledge}.
\begin{table}[!htbp]
\centering
\begin{tabular}{@{}cccc@{}}
\hline
\textbf{Datasets} & \textbf{Training} & \textbf{Validation} & \textbf{Test} \\
\hline \hline
Performance       & 200,000 & 250 & 250 \\
Non-performance   & 200,000 & 250 & 250 \\
\hline
\end{tabular}
\caption{Training, validation and test dataset}
\label{table:sizeOFdatasets}
\end{table}

\paragraph{\textbf{Methodology}}
To implement this approach, we employ Mistral-7B as our teacher model. Mistral-7B, with its 7 billion parameters, has demonstrated high accuracy and robust performance in commit classification (section ~\ref{LLM_intro}), making it an ideal candidate for knowledge transfer. We utilize response-based knowledge distillation, where the teacher model generates predictions for the training data, which are subsequently used to train the student model.

Initially, we fine-tune the teacher model on our commit classification task. For prompting, we follow the approach discussed in Section~\ref{LLM_intro}. Using this prompt, the teacher model labels the 7 million commit messages. To ensure a balanced training set, we randomly select a subset of commits to create 200K performance and 200K non-performance labeled examples. These labels, generated by the teacher model, serve as soft labels, providing probability distributions over the classes rather than hard labels. This probabilistic labeling enhances the student model's ability to generalize, as it learns from the nuanced information provided by the teacher.

Let \( T(x) \) represent the probability distribution over the classes produced by the teacher model for an input \( x \). The soft labels are given by:
\[ T(x) = \left[ P(y = 1 \mid x; \theta_T), P(y = 0 \mid x; \theta_T) \right] \]
where \( \theta_T \) represents the parameters of the teacher model.

We then train our student model, a 125 million parameter BERT, using these soft labels. The student model, denoted as \modelkd{}, is trained to minimize the cross-entropy loss between its predictions and the soft labels provided by the teacher model. The loss function for training the student model is defined as:
\[ L(\theta_S) = -\frac{1}{N} \sum_{i=1}^{N} \left[ T_1(x_i) \log S(x_i) + T_0(x_i) \log (1 - S(x_i)) \right] \]
where \( N \) is the number of training examples, \( T_1(x_i) \) and \( T_0(x_i) \) are the probabilities of the positive and negative classes from the teacher model for input \( x_i \), and \( S(x_i) \) is the probability of the positive class predicted by the student model for the same input \( x_i \).

\begin{figure}[htbp]
    \centering
    \scalebox{0.5}{\input{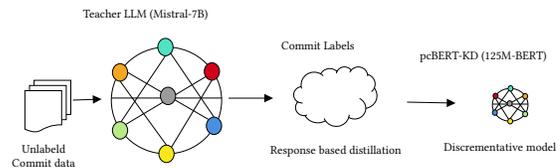}}
    \caption{Overview of knowledge distillation approach}
    \label{fig:KDpipeline}
\end{figure}

This training process enables the student model to capture complex patterns and implicit features that are inherent in the data but might be overlooked by simpler models and keyword-filtering. The resulting student model, referred to as \modelkd{}, benefits from the distilled knowledge of the teacher model, achieving high accuracy while maintaining scalability. Figure~\ref{fig:KDpipeline} illustrates the steps involved in our knowledge distillation approach.


\paragraph{\textbf{Observation}}
Our evaluations demonstrate that \modelkd{} achieves the highest accuracy with an F1-score of 0.93 in classifying performance commits, which is comparable to the F1-score of 0.92 achieved by Mistral-7B.

\subsection{\protect\tool{} Implementation}
\label{sec:miner}
This section outlines the implementation details of \tool{}, the repository mining tool designed to curate a large-scale performance-related commit dataset. \tool{} is built on top of PyDriller~\cite{spadini2018pydriller}, an efficient library for mining software repositories. \tool{} offers various configurable knobs to fine-tune the data collection process. Users can specify parameters such as programming languages, repository star counts, and the number of files or functions impacted by commits. By leveraging the core \modelkd{} for performance commit classification, \tool{} provides a robust mechanism to identify and extract relevant commits. The deduplication step further ensures the quality and uniqueness of the mined dataset.

\paragraph{\textbf{Configuration of \tool{}}}
The data collection process begins by targeting public repositories on GitHub. \tool{} allows users to configure various properties for data collection, including filtering by programming languages and repository star counts. \tool{} further can be configured to collect commits impacting a specified number of files or functions, thereby tailoring the data collection process to specific research needs. For each configured repository, \tool{} crawls the commit history of the \emph{main} branch. Designed to operate in parallel across multiple nodes, \tool{} can mine data concurrently, significantly enhancing the efficiency of the data collection process.

At its core, \tool{} utilizes \modelkd{} to accurately identify performance-related commits. Once such a commit is detected, \tool{} extracts the before and after versions of the source code and the associated metadata.

\paragraph{\textbf{Deduplication}}
\tool{} further implements a data deduplication strategy. For each performance-related commit, \tool{} calculates the MD5 hash of the modified functions (both before and after versions). While mining the commits, \tool{} matches these hashes to ensure a unique list of commits is maintained. Since \tool{}'s data collection process is distributed in nature, it performs an additional deduplication step at the end across concurrent nodes.

\section{Evaluation}
\subsection{RQ3: Evaluating Language Models}
\label{section:RQ3}
In this section, we analyze the accuracy and computational overhead of the proposed language models \textit{\modelkd{}} and \textit{\modelhs{}}. To provide a comprehensive comparison, we also evaluate \textit{Mistral-7B} and its quantized variant, \textit{Mistral-AWQ}. Additionally, we include keyword-filtering as the baseline for our comparative analysis.

\paragraph{\textbf{Accuracy in Detecting Performance Commits}}
To assess the accuracy of the different strategies, we compute precision, recall, F1-score, False Positive Rate (FPR), and accuracy for both performance and non-performance classes. In our evaluation, the F1-score of the performance class is considered the most important metric, as the primary objective of this research is to identify performance-related commits. We conduct our evaluation five times and calculate the mean of all the metrics. We use paired sample t-tests to determine if the observed performance differences between the two strategies are statistically significant, ensuring the p-value is less than 0.05. The results of statistical significance test are included in replication package. Table~\ref{table:performance_metrics} summarizes the accuracy results.

The results show that the \textit{\modelkd{}} model outperforms the other models in terms of F1-score for the performance class. With a precision of 0.90, a recall of 0.97, and an F1-score of 0.93 for the performance class, \textit{\modelkd{}} demonstrates a strong ability to accurately identify performance-related commits. Additionally, its low false positive rate (FPR) of 0.10 further highlights its reliability.

Comparatively, the \textit{\modelhs{}} model shows a significant drop in performance, with an F1-score of 0.75 for the performance class. Although it has a reasonable precision and recall for performance (0.77 and 0.74 respectively), its higher FPR of 0.21 suggests a greater likelihood of false positives.

The Mistral models, particularly \textit{Mistral-7B} and \textit{Mistral-AWQ}, also show strong classification performance but slightly less optimal than \textit{\modelkd{}}. The superior performance of \textit{\modelkd{}} is due to knowledge distillation, enhancing generalization and efficiency~\cite{hinton2015distilling}. 

The baseline, \textit{Keyword-Filtering}, lags significantly behind with an F1-score of 0.59 for the performance class, demonstrating the clear advantage of the language models evaluated in this study.

\begin{table*}[!htbp]
\centering
\small
\begin{tblr}{
  row{2} = {c},
  row{3} = {c},
  row{6} = {c},
  row{9} = {c},
  column{2} = {c},
  cell{1}{1} = {r=2}{},
  cell{1}{2} = {r=2}{},
  cell{1}{3} = {c=4}{c},
  cell{1}{7} = {c=4}{c},
  cell{3}{1} = {c=10}{},
  cell{4}{3} = {c},
  cell{4}{4} = {c},
  cell{4}{5} = {c},
  cell{4}{6} = {c},
  cell{4}{7} = {c},
  cell{4}{8} = {c},
  cell{4}{9} = {c},
  cell{4}{10} = {c},
  cell{5}{3} = {c},
  cell{5}{4} = {c},
  cell{5}{5} = {c},
  cell{5}{6} = {c},
  cell{5}{7} = {c},
  cell{5}{8} = {c},
  cell{5}{9} = {c},
  cell{5}{10} = {c},
  cell{6}{1} = {c=10}{},
  cell{7}{3} = {c},
  cell{7}{4} = {c},
  cell{7}{5} = {c},
  cell{7}{6} = {c},
  cell{7}{7} = {c},
  cell{7}{8} = {c},
  cell{7}{9} = {c},
  cell{7}{10} = {c},
  cell{8}{3} = {c},
  cell{8}{4} = {c},
  cell{8}{5} = {c},
  cell{8}{6} = {c},
  cell{8}{7} = {c},
  cell{8}{8} = {c},
  cell{8}{9} = {c},
  cell{8}{10} = {c},
  cell{9}{1} = {c=10}{},
  cell{10}{3} = {c},
  cell{10}{4} = {c},
  cell{10}{5} = {c},
  cell{10}{6} = {c},
  cell{10}{7} = {c},
  cell{10}{8} = {c},
  cell{10}{9} = {c},
  cell{10}{10} = {c},
  hline{1,3-11} = {-}{},
  hline{2} = {3-10}{},
}
\textbf{Model}          & \textbf{Accuracy} & \textbf{Performance} &                  &                   &                  & \textbf{Non-Performance} &                  &                   &                  \\
                        &                   & \textbf{Precision}   & \textbf{Recall}  & \textbf{F1 Score} & \textbf{FPR}     & \textbf{Precision}       & \textbf{Recall}  & \textbf{F1 Score} & \textbf{FPR}     \\ \hline
\textbf{125M-parameter BERT Models} &                   &                      &                  &                   &                  &                          &                  &                   &                  \\
\textit{\modelkd{}}              & \textbf{0.93}           & 0.90             & \textbf{0.97}          & \textbf{0.93}           & 0.10           & \textbf{0.97}                  & 0.90          & \textbf{0.93}           & 0.03           \\
\textit{\modelhs{}}              & 0.77           & 0.77             & 0.74          & 0.75           & 0.21           & 0.76                  & 0.79          & 0.78           & 0.26           \\
\textbf{7B-parameter Large Language Models} &                   &                      &                  &                   &                  &                          &                  &                   &                  \\
\textit{Mistral-7B}              & 0.92           & 0.93             & 0.90          & 0.92           & 0.06           & 0.91                  & 0.94          & 0.92           & 0.10           \\
\textit{Mistral-AWQ}             & 0.91           & 0.95             & 0.85          & 0.90           & 0.04           & 0.87                  & 0.96          & 0.91           & 0.15           \\
\textbf{Baseline}       &                   &                      &                  &                   &                  &                          &                  &                   &                  \\
\textit{Key-Filtering}           & 0.67           & 0.75             & 0.48          & 0.59           & 0.15           & 0.63                  & 0.84          & 0.72           & 0.51          
\end{tblr}
\caption{Performance Metrics of Various Models}
\label{table:performance_metrics}
\end{table*}

\paragraph{\textbf{Computational Performance}} 
We evaluate the computational performance of various strategies on three contemporary CPU and GPU architectures: AMD Threadripper Pro 5955WX (CPU), NVIDIA RTX 4090 (two consumer-level GPU nodes), and NVIDIA A10 (one GPU node on Oracle Cloud). Table~\ref{table:hardware_spec} details the hardware specifications of these architectures. To assess the quantized Mistral 7B model, we use Mistral-AWQ for GPUs and Mistral-GGUF for the CPU. As keyword-filtering runs only on the CPU, we do not test the approach on GPU nodes. We measure computational performance in terms of throughput (tokens per second), and for keyword-filtering, which does not rely on tokens, we calculate the equivalent tokens of the ground truth dataset for comparison with other language models. We run each test five times, and calculate the mean throughput. We further perform statistical significance testing by conducting a t-test to ensure a p-value less than 0.05. The results of the statistical t-test are reported in the replication package. 

\begin{figure}[!htbp]
\centering
    \includegraphics[width=0.99\linewidth]{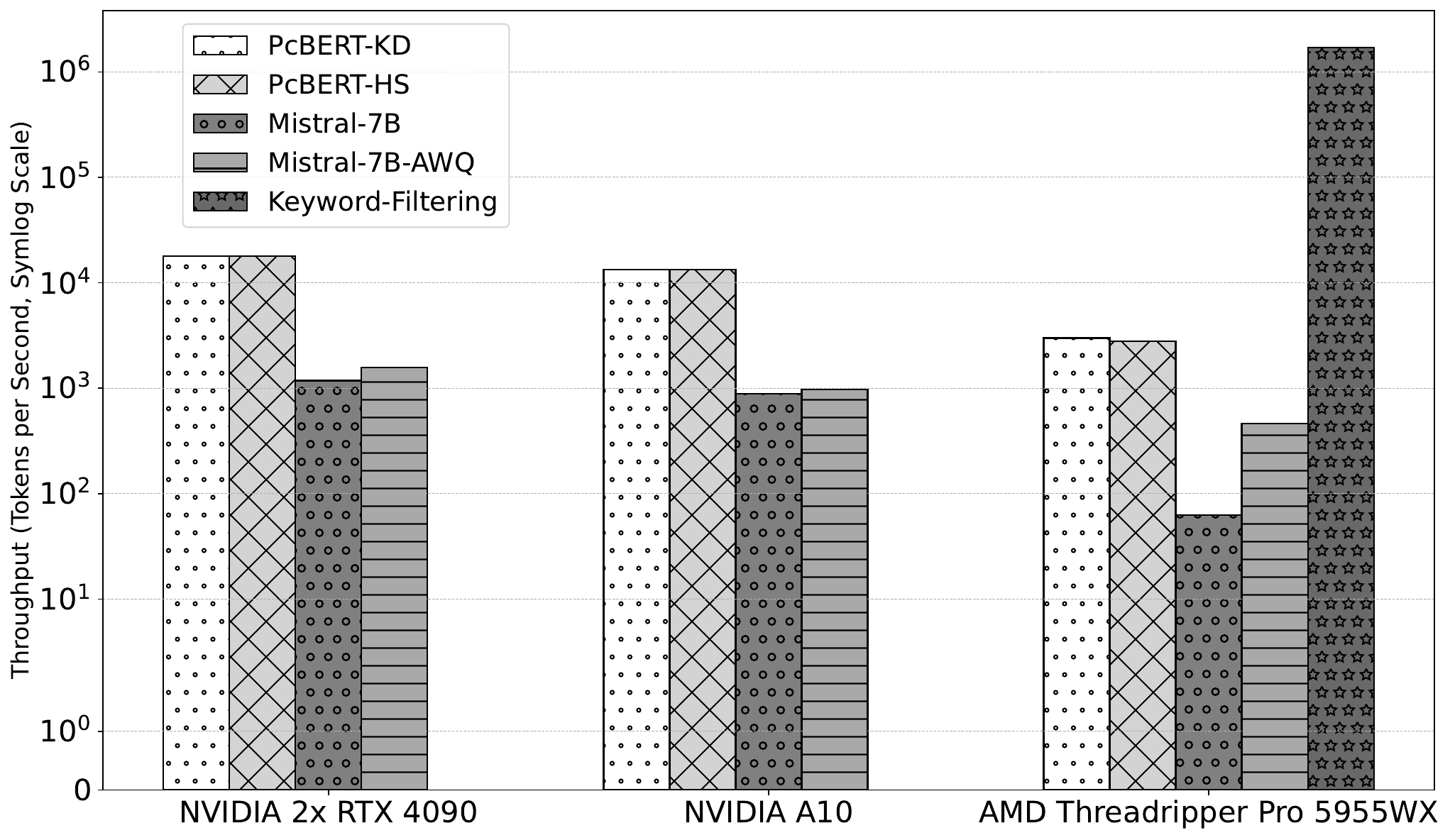}
    \caption{Model inference throughput (Tokens per second) on CPU and GPUs plotted at symlog scale.}
    \label{fig:Throughput}
\end{figure}

The computational performance is plotted on figure~\ref{fig:Throughput}. Based on the evaluation we make following key observations:

\begin{itemize}
    \item Both of the 125M-parameter BERT model (\textit{\modelkd{}} and \textit{\modelhs{}}) consistently demonstrate the highest throughput across different hardware configurations, significantly outperforming the Mistral variants.
    \item \textit{\modelkd{}} achieves 2,989.57 tokens per second on a CPU, surpassing the Mistral models, with Mistral-7B reaching only 79.81 tokens per second. \modelkd{}'s superior performance on CPUs makes it a practical choice for real-world deployment.
    \item While results demonstrate that Mistral-AWQ is more efficient than Mistral-7B, indicating that optimization by quantization enhances throughput, both models lag significantly behind the \modelkd{} variants.
\end{itemize}

\begin{table*}[!htbp]
\centering
\small
\begin{tabular}{@{}cccccc@{}}
\hline
\textbf{Unit} & \textbf{Model} & \textbf{Architecture} & \textbf{Processing Units} & \textbf{On-Chip Memory} & \textbf{Off-Chip Memory} \\ \hline \hline
GPU & NVIDIA 2x RTX 4090 & Ada Lovelace & 256 SMs   & 128 KB L1 per SM, 72 MB L2 &  24 GB GDDR6X per GPU \\ 
GPU & NVIDIA A10 & Ampere  & 72 SMs   & 128 KB L1 per SM, 40 MB L2 &  24 GB GDDR6\\ 
CPU & AMD Threadripper Pro 5955WX & Zen 3 & 16 Cores   & 32 KB L1, 512 KB L2, 32 MB L3 &  128 GB \\ \hline
\end{tabular}
\caption{Hardware specifications of evaluated architectures}
\label{table:hardware_spec}
\end{table*}

\subsection{\protect\tool{} Evaluations Settings}
We configure \tool{} to collect performance-related commits from GitHub repositories written in three programming languages: Python, C++, and Java. To ensure the collection of high-quality and diverse performance commits, we target public repositories with a star count of $>=20$ for each programming language. This criterion resulted in a total of 322K repositories. Following prior work~\cite{mahbub_explaining_2023}, we filter out commits that involve changes to more than one function during the mining process to avoid tangled commits.

To efficiently mine these repositories, we deploy \tool{} on a CloudLab~\cite{Duplyakin+:ATC19} cluster. The cluster consists of 50 nodes, each equipped with Intel® Xeon® D-1548 processors running at 2.00GHz and 64GB of DDR4 RAM. This robust infrastructure allows \tool{} to operate in parallel across multiple nodes, enabling the concurrent mining of data.

\begin{table}[!htbp]
\centering
\resizebox{\columnwidth}{!}{%
\begin{tabular}{ccccc}
\hline
\textbf{Language} & \textbf{\# Repos} & \textbf{\# Commits} & \textbf{\# size}  & \textbf{\# Perf Bugs} \\
\hline \hline
Python & 190K & 59M   & 6.8Gb & 114K   \\   %
C++     & 59K & 57M  & 48Gb  & 217.9K \\  
Java & 73K & 54M  & 4.8Gb & 76.6K  \\    

\hline
\end{tabular}%
}
\caption{Statistics of mined performance bug dataset.}
\label{table:minnedDataset}
\end{table}

\paragraph{\textbf{Result summary}} Table~\ref{table:minnedDataset} summarizes the mining result. \tool{} mined 170M commits across Python, C++, and Java. From this extensive collection, we identified a substantial number of performance-related commits: 114K in Python, 217.9K in C++, and 76.6K in Java, totaling over 408.5K performance commits. The significant number of collected performance commits underscores the importance of developing efficient tools to identify and address performance bugs across different programming languages.

\subsection{RQ4: Evaluating Mined Dataset}
\label{subsec:eval_mined_data}
\label{evalDataset}
\subsubsection{\textbf{Manual Validation}}
To ensure the accuracy of the mined performance commits, three authors independently reviewed and analyzed a statistically significant sample of 384 commits, chosen to meet a 95\% confidence level with a ±5\% confidence interval. This thorough examination aimed to validate whether the commits were genuinely performance-related. During this process, any disagreements were identified and resolved through discussion and consensus. Ultimately, the authors concluded that 96\% of the identified performance commits were true positives. This rigorous validation process confirmed a high level of reliability and accuracy in classifying performance commits.

\subsubsection{\textbf{Analysis of Mined Dataset}}

\begin{table*}[!htbp]
\centering
\label{table:categoryDataset1}
\begin{adjustbox}{max width=\linewidth}
\begin{small}
\begin{tabular}{c l c c c}
\hline
\textbf{Category} & \textbf{Sub-category} & \textbf{Python} & \textbf{C++} & \textbf{Java} \\ \hline \hline

\multirow{4}{*}{\textbf{API Misuse}} & Incorrect API Usage & 697 & 1040 & 632 \\ \cline{2-5}
& Deprecated API & 184 & 156 & 80 \\ \cline{2-5}
& Redundant API Calls & 2616 & 2430 & 1503 \\ \cline{2-5}
& Misc. API Misuse & 9 & 34 & 20 \\ \hline

\multirow{3}{*}{\textbf{Memory Inefficiency}} & Memory Leak & 104 & 805 & 106 \\ \cline{2-5}
& Unnecessary Memory Allocation & 6380 & 63486 & 5894 \\ \cline{2-5}
& Misc. Memory Inefficiency & 740 & 4221 & 227 \\ \hline

\multirow{4}{*}{\textbf{Poor Concurrency Control}} & Thread Contention & 1037 & 1933 & 1082 \\ \cline{2-5}
& Unnecessary locks & 1539 & 4138 & 674 \\ \cline{2-5}
& Unnecessary Thread Synchronization & 7869 & 15614 & 10619 \\ \cline{2-5}
& Misc. Poor Concurrency Control & 463 & 1131 & 232 \\ \hline

\multirow{4}{*}{\textbf{Inefficient I/O}} & Inefficient Disk I/O & 9568 & 4802 & 5392 \\ \cline{2-5}
& Inefficient Caching & 1057 & 443 & 700 \\ \cline{2-5}
& Unnecessary Logging & 6656 & 4071 & 3717 \\ \cline{2-5}
& Misc. Inefficient I/O & 3093 & 1899 & 1389 \\ \hline

\multirow{3}{*}{\textbf{Network Bottlenecks}} & Inefficient Data Transfer & 690 & 602 & 503 \\ \cline{2-5}
& Excessive Network Calls & 1359 & 620 & 736 \\ \cline{2-5}
& Misc. Network Bottlenecks & 81 & 89 & 44 \\ \hline

\multirow{6}{*}{\textbf{Inefficient Algorithm/Data-structure}} & Suboptimal Data Structures & 7644 & 10182 & 4612 \\ \cline{2-5}
& Suboptimal Algorithm & 3089 & 4594 & 1610 \\ \cline{2-5}
& Expensive Operation & 14490 & 21979 & 7916 \\ \cline{2-5}
& Unnecessary computations & 24360 & 25187 & 19555 \\ \cline{2-5}
& Inefficient Loops & 3590 & 5162 & 2063 \\ \cline{2-5}
& Misc. Inefficient Algorithm/Data-structure & 9253 & 11082 & 3922 \\ \hline

\multirow{3}{*}{\textbf{Parallelization}} & Missing Parallelism & 1899 & 3014 & 477 \\ \cline{2-5}
& Inefficient Parallelism & 471 & 1347 & 212 \\ \cline{2-5}
& Misc. Parallelization & 157 & 348 & 37 \\ \hline

\multirow{3}{*}{\textbf{Micro-architectural}} & Data Locality & 62 & 1880 & 33 \\ \cline{2-5}
& Missed Compiler Optimization & 74 & 3486 & 37 \\ \cline{2-5}
& Misc. Micro-architectural & 97 & 849 & 7 \\ \hline

\multirow{1}{*}{\textbf{Other}} & Misc. Other & 4681 & 21394 & 2562 \\ \hline

\textbf{Total} & & \textbf{114009} & \textbf{217918} & \textbf{76593} \\ \hline

\end{tabular}
\end{small}
\end{adjustbox}
\caption{Frequency counts of performance bug categories for Python, C++, and Java.}
\label{table:categoryDataset1}
\end{table*}

\paragraph{\textbf{Objective}}
To understand whether the collected dataset  covers a wide range of performance bugs, the next step in our pipeline is to investigate RQ4 by examining the distribution of performance commits across various performance pitfalls. 

\paragraph{\textbf{Methodology}}

We leverage existing performance bug taxonomies to define a two-level hierarchical classification system. We further refine this classification based on manual observations. While this finer-grained categorization helps understand the dataset’s diversity, manually mapping labels for large-scale datasets is impractical, necessitating an automated approach. To address this, we employ an LLM-based method to automate the task efficiently.

Recent work~\cite{colavito_leveraging_2024,he_annollm_2024} shows that LLMs can effectively perform such tasks with sufficient guidance, achieving comparable performance to human labeling. Hierarchical categorization involves complex, multi-step reasoning and more context than binary classification. Our dataset includes rich contextual information, like source code changes related to performance issues. By leveraging this, we use the \emph{chain-of-thoughts (CoT)}~\cite{fan_chain--thought_2023} prompting method, which enhances LLM performance in zero-shot settings by guiding step-by-step reasoning~\cite{kojima2023large}. Our experimental setup focuses on zero-shot settings to understand the distribution of performance issues.

\begin{table}[!t]
\centering
\begin{tabular}{p{0.95\columnwidth}}
\hline
\textbf{\textcolor{brown}{"prompt"}}: You are provided with a GitHub commit in the following format:
\\ \texttt{\{commit\_message\}}
\texttt{\{original\_code\}}
\texttt{\{modified\_code\}}
\texttt{\{code\_diff\}}\\
\textbf{Task description:} Task description is provided along with the reasoning steps\\
\textbf{Category description:} Category/sub-category description is provided \\
\textbf{Output description}: Output instructions are provided to get model output in the desired format \\
\textbf{\textcolor{brown}{"output"}}: \{category label/s\}\\
\hline
\end{tabular}
\caption{CoT Prompt template for categorization of performance bug types (Please refer to replication package for the full prompt used in the study)}
\label{tab:promptTemplate2}
\end{table}

\paragraph{\textbf{Prompt Design}} We adopt a prompt design approach similar to that in section~\ref{LLM_intro}, integrating the CoT technique as per existing literature~\cite{wei_chain--thought_2023}. Table~\ref{tab:promptTemplate2} displays the CoT prompt used for performance commit categorization. Our experimental settings provide the model with context input, task description, and category definitions. For root cause analysis of performance issues identified in the first pipeline phase, we include detailed context information: the \emph{commit message, before and after source code of the modified method, and code diff}. The commit message highlights the targeted performance issues and rationale for changes. The \emph{code diff} shows altered lines, the \emph{original code} offers context and identifies prior inefficiencies, and the modified code reveals performance optimizations. The task description outlines the task’s definition and hierarchical reasoning steps. Category definitions clarify each category, aiding the model in accurate categorization based on the root cause.
\paragraph{\textbf{Observation}}
The resulting categorization of the performance-related commits is presented in Table~\ref{table:categoryDataset1}. Three authors manually analyzed the categorization to ensure its accuracy.

The distribution of performance commits across categories identified by \modelkd{} demonstrates its proficiency in detecting diverse performance bug fix commits. The model effectively categorizes performance issues across multiple programming languages, validating its robustness and versatility.

To identify the most prevalent and common performance bugs across languages, we calculated a significance metric, \(\sigma\), based on the categorization data. The significance metric is determined by the ratio of the number of commits in a specific classification category to the total number of commits in the respective language. This metric highlights the relative importance of each category within the context of the language. The metric is calculated using the following equation:
\[
\sigma = \frac{\text{Number of commits in category}}{\text{Total number of commits in language}}
\]
Table~\ref{table:categoryDataset2} presents a comparison of classification categories across different programming languages, sorted based on their significance metric, \(\sigma\). Based on the table we make following observations:

\begin{itemize}
    \item \textbf{Unnecessary Computations Across Languages:} \textit{Unnecessary computations} are a prevalent issue across all programming languages: Python (\(\sigma = 0.21\)), C++ (\(\sigma = 0.12\)), and Java (\(\sigma = 0.26\)). This highlights a widespread problem with inefficient coding practices leading to redundant calculations. \emph{\textbf{This suggests a common need for developers to focus on optimizing computational logic and reducing unnecessary operations.}}
    \item \textbf{High Memory Allocation in C++:} In C++, \textit{Unnecessary Memory Allocation} exhibits a high significance metric (\(\sigma = 0.29\)), surpassing all other categories within the same language. This highlights a major issue with memory management in C++, due to its manual memory handling compared to the automatic garbage collection in Python and Java. \emph{\textbf{The large number of performance bug-fix-related commits addressing memory allocation inefficiencies in C++ highlights the importance of this area, suggesting a need for improved memory management tools or practices.}}
    \item \textbf{High Unnecessary Thread Synchronization in Java:} The significance metric for \textit{Unnecessary Thread Synchronization} is substantial in Java (\(\sigma = 0.14\)), highlighting concurrency management as a notable performance bottleneck in this language. \textbf{Effective concurrency control mechanisms or tools are crucial for optimizing performance in Java.}
\end{itemize}

\begin{table}[t]
\centering

\begin{adjustbox}{max width=\linewidth}
\begin{small}
\begin{tabular}{l l c}
\toprule
\textbf{Language} & \textbf{Classification Category} & \textbf{Significance Metric (\(\sigma\))} \\ \midrule
\hline
\multirow{5}{*}{\textbf{Python}} 
& \cellcolor{red!30}Unnecessary Computations & 0.21 \\ 
& \cellcolor{blue!30}Expensive Operations & 0.13 \\ 
& \cellcolor{yellow!30}Inefficient Disk I/O & 0.08 \\ 
& \cellcolor{orange!30}Misc. Inefficient Algorithm/Data-structure & 0.08 \\ 
& \cellcolor{green!30}Unnecessary Thread Synchronization & 0.07 \\ \midrule

\multirow{5}{*}{\textbf{C++}} 
& Unnecessary Memory Allocation & 0.29 \\ 
& \cellcolor{red!30}Unnecessary Computations & 0.12 \\ 
& \cellcolor{blue!30}Expensive Operations & 0.10 \\ 
& \cellcolor{green!30}Unnecessary Thread Synchronization & 0.07 \\ 
& \cellcolor{orange!30}Misc. Inefficient Algorithm/Data-structure & 0.05 \\ \midrule

\multirow{5}{*}{\textbf{Java}} 
& \cellcolor{red!30}Unnecessary Computations & 0.26 \\ 
& \cellcolor{green!30}Unnecessary Thread Synchronization & 0.14 \\ 
& \cellcolor{blue!30}Expensive Operations & 0.10 \\ 
& \cellcolor{yellow!30}Inefficient Disk I/O & 0.07 \\ 
& \cellcolor{cyan!30}Suboptimal Data Structures & 0.06 \\ \bottomrule

\end{tabular}
\end{small}
\end{adjustbox}
\caption{Top 5 performance bug classification categories across languages, sorted by significance metric (\(\sigma\)). The significance metric is calculated as the ratio of the number of commits in a specific classification category to the total number of commits in the respective language.}
\label{table:categoryDataset2}
\end{table}

\subsection{RQ5: Data Usability Analysis} 
\label{sec:eval_usability}

This research work aims to mine performance bug commits at scale. However, measuring the usability of this data is non-trivial and may require using the data in real-world scenarios. Motivated by prior research~\cite{ding2024vulnerability,chen2023diversevul}, we apply our approach-generated data to LLMAPIDet~\cite{wei_demystifying_2024}, a data-centric tool for detecting API misuse, to measure its usability by detecting performance-related API misuse. 

\paragraph{\textbf{LLMAPIDet's Approach}}LLMAPIDet uses large language models (LLMs) for deep learning-based API misuse detection and patching. ChatGPT enriches API misuse rules with manually created examples, including code before and after fixes, and defined rules. Three authors labeled 4,224 commits, with each analysis averaging 6.3 minutes. Each author spent 443.53 hours, totaling 1,330.56 man-hours. Manual labeling identified 891 confirmed API misuses to construct the knowledge base. With distilled knowledge, LLMAPIDet generates code explanations for API uses. The paper reported a 32.33\% precision of the proposed approach.

\paragraph{\textbf{Study goal}}
In this study, we replicate the process of LLMAPIDet with our approach-generated API misuse data rather than manually labeled data. The goal is to demonstrate that the \tool dataset can generate knowledge equivalent to what LLMAPIDet authors produced manually. We aim to show that leveraging the scaled data collected by \tool, we can enrich the knowledge database and achieve higher performance without manual labor.

\paragraph{\textbf{Study approach}}
We perform an analysis on performance API-misuse categories as our earlier evaluations suggest that API misuse is one of the top reasons for performance bugs, and LLMAPIDet is also designed for API-misuse detection. As discussed in~\autoref{subsec:eval_mined_data}, our proposed approach has categorized 3,201 performance bug commits related to API misuses in Python.

To analyze LLMAPIDet performance with different data sizes, we construct a knowledge base with five dataset points (500, 1000, 1500, 2000, 2500, and 3000 data points). Each subset is sampled five times with replacement for statistical reliability. This rigor is crucial for observing the relationship between dataset size and detection performance.

\paragraph{\textbf{Observation}}We evaluate LLMAPIDet performance for 111 performance API misuse ground truth cases using the knowledge base. Table~\ref{table:db_accuracy_recall} shows the accuracy of LLMAPIDet with different knowledge base sizes. The performance trend supports the hypothesis that data volume enhances the model's learning capability and detection accuracy. For instance, accuracy improves from 35.14\% at 500 commits to 45.05\% at 3000 commits, showing significant scalability benefits. Moreover, our approach saves the time and effort of manual labeling. These insights are crucial for future optimizations in API misuse and other software fault types.

\begin{table}[h]
\centering
\small
\vspace{-0.3cm}
\resizebox{0.5\columnwidth}{!}{%
\begin{tabular}{cc}
\hline
\textbf{Dataset Size} & \textbf{Accuracy} \\
\hline \hline
500 & 35.14\% \\
1000 & 37.30\% \\
1500 & 38.92\% \\
2000 & 41.62\% \\
2500 & 43.42\% \\
3000 & 45.05\% \\
\hline
\end{tabular}
}
\caption{\texttt{LLMAPIDet} accuracy of detecting performance bugs related to API misuse using knowledge bases formed from different dataset sizes.}
\label{table:db_accuracy_recall}
\end{table}
\vspace{-0.4cm}


\section{Threats to validity}
\label{sec:threats}
\vspace{-0.1cm}
\paragraph{\textbf{Construct validity}}
Construct validity is compromised if the language models inaccurately interpret commit messages due to incomplete heuristics (\modelhs), over-reliance on the imprecise teacher model (\modelkd), or an inability to learn the nuances of performance-related commits. To ensure robust construct validity, we manually examined the mined commit dataset to verify that the commits are genuinely related to performance-bug fixes.    
\paragraph{\textbf{Internal validity}}
Internal validity is threatened by selection biases that might influence the study's outcomes. If the selected commits are not representative of typical performance issues, it may lead to biased findings. To mitigate these threats, we employed random sampling on a large corpus of data during model training, ensuring unbiased sample of commits.
\paragraph{\textbf{External validity}} External validity concerns the generalizability of the study’s findings. To enhance external validity, we included diverse samples from various repositories and programming languages. Additionally, validating the mined dataset on the NLP tool, LLMAPIDet, to detect API misuse related to performance bugs ensured that the findings are broadly applicable and practical, thereby strengthening the overall applicability and impact of the research outcomes.
\vspace{-0.2cm}
\section{Conclusions}
\label{sec:conclusion}
\vspace{-0.2cm}
In this paper, we present \tool{}, a repository mining tool designed to identify performance-related bug-fix commits at scale. To classify a commit as performance bug-fix related with high accuracy and low computational overhead, we propose two transformer models: \modelkd{} and \modelhs{}. Through empirical evaluation, we demonstrate that \modelkd{} excels in performance commit detection, achieving an F1-score of 0.93 compared to the baseline keyword-filtering approach, while offering a 37.5$\times$ speedup over large language models such as Mistral-7B. Utilizing \modelkd{} as a performance commit detector, \tool{} identified a total of 408.5K performance-related bug-fix commits across Python, C++, and Java. We performed an analysis of the collected dataset, providing insights to guide future performance engineering research. Additionally, we show that the significant dataset size further improves the accuracy of data-centric performance detection techniques by 28\%.

\bibliographystyle{ACM-Reference-Format}
\bibliography{currentproject}










\end{document}